# Towards Sealed Resistive Plate Chambers

**L. Lopes**[a,1]**, P. Assis**[b,c]**, A. Blanco**[a]**, P. Fonte**[a,d]**, and M. Pimenta**[b,c]

[a] *Laboratório de Instrumentação e Física Experimental de Partículas (Lip),*
  *Departamento de Física da Universidade de Coimbra, 3004-516 Coimbra, Portugal*

[b] *Laboratório de Instrumentação e Física Experimental de Partículas (Lip),*
  *Av. Professor Gama Pinto, n. 2, Complexo Interdisciplinar (3is), 1649-003 Lisboa, Portugal*

[c] *Departamento de Física, Instituto Superior Técnico, Universidade de Lisboa*
  *Avenida Rovisco Pais, n. 1, 1049-001 Lisboa, Portugal*

[d] *Coimbra Polytechnic – ISEC, Coimbra, Portugal*

  *E-mail*: luisalberto@coimbra.lip.pt

Abstract: The mitigation of human-induced climate change is of crucial importance for the sustainability of humankind. For this aim the RPC community has exerted considerable effort over the last decade to reduce the emission of greenhouse gases from our detectors. These included searching for new eco-friendly gases, implementing recovery and/or recirculation systems, improving gastightness and using new materials and approaches in detector conception and operation for the reduction of gas flow rates.

Along this line of work, we present here an RPC architecture aimed at a dramatic reduction of gas use in chambers meant for low-rate operation. Two chambers were tested for more than six months with zero gas flow showing no evidence of time-related effects, allowing to consider that permanently sealed RPCs may be within reach, with obvious practical and environmental advantages.

---

[1] Corresponding author.

**Contents**

**1. Introduction**

Resistive Plate Chambers (RPCs) [1] can be found in a large number of experiments, mainly in indoor but also in outdoor environments [2-6]. In all these experiments/applications Hidrofluorocarbons (HFCs) are present, in some cases constituting 100 % of the gas mixture. In recent years the environmental problems and consequent constraints related to the use of such gases caused concern in the RPC Community. Many [7-10] studies searching for replacement gases with acceptable Global Warming Potential (GWP) have shown that it is possible to reduce the amounts of HFCs in the mixtures but unfortunately it seems impossible to achieve the needed performances, at least when very good timing is required. For indoor applications, where it is possible to use recirculation and recycling [11], the gas consumption rate could be reduced, but unfortunately this could not be implemented in outdoor applications where the stations are very far from each other. Mostly considering outdoor applications, some systematic studies were done in the laboratory and outdoors (in harsh field conditions) [12-13], operating the RPCs at very low gas flow rates while using very simple and low-cost gas systems. The results were encouraging, being possible to operate at gas flow rates below 4 cc.min$^{-1}$ for long periods of time. However more data and time are needed to prove the stability/applicability of this design. At the moment it seems impossible to survive without $C_2H_2F_4$ in any application, and for timing $SF_6$ is a requirement to operate at higher gains with low streamer fraction reaching for the last picoseconds.

Complementing other lines of work, we decided to invest our efforts in the development of sealed RPCs. This would allow to continue using the optimal gas mixtures while considerably reducing the cost in gas consumption, distribution, recovery and recycling systems. Indeed, for some applications the independence from a permanent gas supply would constitute a paradigm shift in the applicability of RPCs.

## 2. The detector

It was clear since the beginning that the development for trigger-like applications would be an easier task than for timing, mainly because of the gap width. All the work presented here was done with 1 mm gas gaps, only float glass was considered as electrode and sensitive areas not smaller than 40x40 cm$^2$.

The objective is to construct a detector with 40x40 cm$^2$ area and between 3 and 7 mm high depending on electrode thickness. Then clean as much as possible the inner surfaces, fill the volume with counting gas and seal it. Very important during material selection is the out-gassing, which should be ideally zero, or the gas purity will become compromised very fast and render the detector impossible to use. Therefore, glues and epoxies should be avoided as much as possible and is mandatory to reduce to zero their contact with the gas. For the resistive electrode float glass is the first and most probably the only possible choice. Very good surface quality (high flatness), high rigidity for detector and gap uniformity, "zero" out-gassing, low thermal expansion coefficient and, most important, very well know inside the RPC Community.

After some not very successful attempts concerning assembly and sealing, **Figure 1** shows the design of the sealed RPCs. The glass sheets are in a first step glued to the copper surface of the external printed circuit board (PCB). It was estimated a maximum epoxy thickness of 0.2 mm. One of these boards has 16 pads of 95x95 mm$^2$ and the other is a single pad of 395x395 mm$^2$. In the end we just get a total volume with 450x450x8 mm$^3$ where 400x400x1 mm$^3$ are expected to be sensitive. In the centre of each pad a small hole was open to weld a cable for signal readout and apply the high voltage. At the beginning all the 16 pads were connected to the same point for quality control. Before start flushing the chamber with Argon, a set of temperature, humidity and absolute pressure sensors was placed around the chamber for precise monitoring of the ambient variables that could influence the chamber behaviour. The sensors and the high voltage power supply were connected to a Raspberry PI microcomputer. All variables are recorded once a minute. The gas distribution is prepared to be easily changed from Argon to Tetrafluoroethane.

These were the only two gases used until now; in the future a mixture of Tetrafluoroethane and Hexafluoride will be also tested. Two equal chambers were built and used during the tests presented and discussed in the next section.

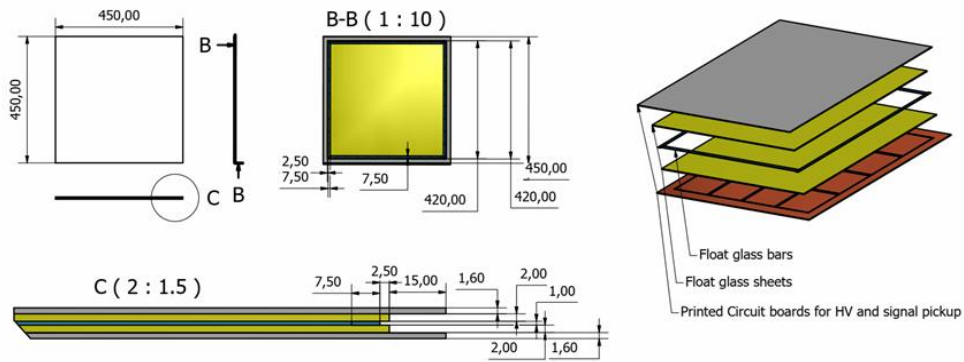

**Figure 1:** Drawings of the sealed RPC. Very simple concept: 2 mm thick glass sheets glued to 1 mm thick glass bars along the sides. Printed circuit boards were glued to the outer surfaces for high voltage supply and signal pickup.

## 3. Results

Since we are developing something new, we are not concerned with the full functionality as a particle detector but rather in the first place in assessing its stability. Thus, the study presented here is based on monitoring the background current over time, the response to sources of irradiation and only after ensuring the chamber's stability we move on to preliminary measurements of some practical quantities. The same tests were done in both chambers and in the end, we kept one for an "endurance" test and use the other to measure some practical quantities.

### 3.1 Argon discharge at continuous gas flow

The first step is always flushing the chambers with Argon for some days with high voltage off. After some days we start to increase the high voltage until establishing a permanent discharge over the full gap area. This is very important to get a better cleanness of the gap surfaces, that way some possible "warm" points could be eliminated or reduced, increasing the chamber lifetime. It is known that dust particles, grease and anything that reduces the gap width will become hot spots for strong discharges and consequent polymerization when complex gases are present. If the gap has clean smooth surfaces, after establishing a permanent discharge the observed current should be stable over time without "jumps" in the short term and be correlated with temperature in the long therm. This long-term behaviour comes from the glass resistivity correlation with temperature [14]. When a permanent discharge over the gap is established any increase in the high voltage will be reflected in the drawn current, respecting Ohm's law if considering normal pressure and temperature (NPT), or at least "constant" temperature since short term (few hours) variations in pressure will have little or no effect on this measurement. This way it should be possible to extract the resistance seen by the current and "precisely" calculate the volume resistivity of the glass used in the gap construction. **Figure 2** shows an example of the tests done with Argon. In the two uppermost panels is shown the current over time and in the lowermost panel the linear dependence from where the volume resistivity is determined. This is for the first chamber built, but the same is observed in the other. The "good health" of the chamber is clear.

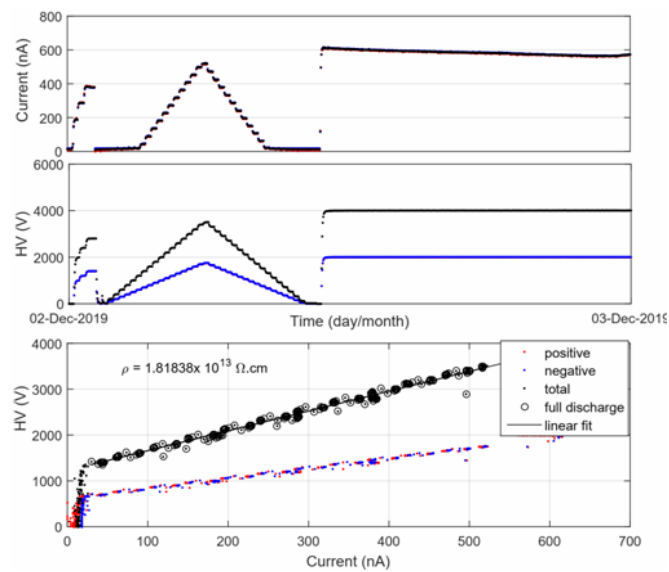

**Figure 2:** Argon discharge quality test. In the uppermost plot the current over time, very smooth. In the middle the high voltage over time. In the lowermost the linear dependence at full gap discharge, used to extract de volume resistivity of the glass used in the gap construction.

### 3.2 Current over time with Tetrafluoroethane

While the high voltage was kept low Tetrafluoroethane was flushed into the chamber. After some days the high voltage was increased until the chamber started to become sensitive to radiation

– 4 –

sources. It was decided to use 5.5 kV (220 Td) as the reference voltage to seal the chamber. After a few days with stable operation the chamber was sealed the 13th of December 2019. In **Figure 3** are shown all the collected variables vs. the time since the chamber was sealed. After starting the operation at a stable reduced electric field (E/N) of 235 Td, by automatically correcting the high voltage for temperature and absolute pressure variations, the long-term behaviour of the current is defined by the temperature and the short term by the absolute pressure [6]. After six months of continuous operation it is apparent that the background current shows no overall trend, besides some temperature and pressure correlation, suggesting a stable gas quality.

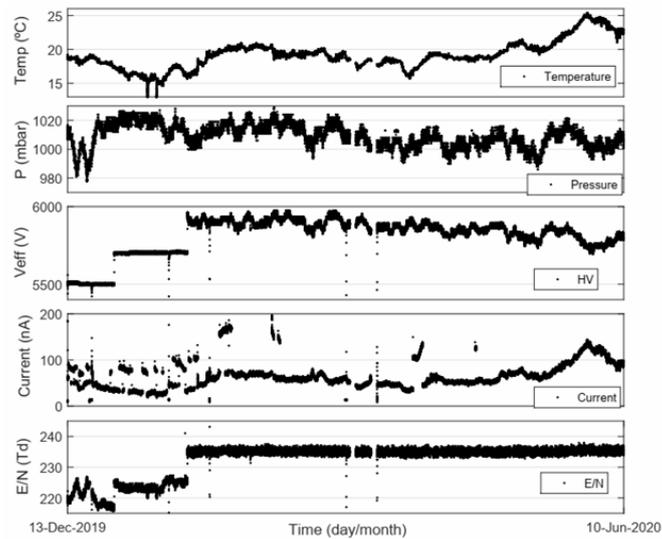

**Figure 3:** Plot of all the variables acquired in the basic monitoring of one sealed RPC as a function of the time since sealing. After a couple of months, the automatic high voltage adjustment was started and the chamber operated at a constant reduced electric field (E/N). In the current plot (4$^{th}$ panel) the sudden spikes indicate the occasions when the chamber was irradiated with a radioactive source for monitoring purposes. The distance between radiation source and chamber was not fixed, which causes different currents for apparently similar conditions.

This observation is further developed in **Figure 4**. The E/N-stabilised background current is (middle panel) represented as a function of the temperature, showing a very good correlation without any evidence of a secondary dependence on the elapsed time. This indicates a stable gas quality over the approx. 5 months period analysed. In the lower panel, data for each 1ºC temperature interval is represented as a function of E/N, showing a weak residual correlation. One may wonder why, being E/N stabilized and without any perceived effect (so the gas gain should be constant), there is still a correlation with temperature. Several effects, difficult to disentangle, that contribute to the observed dark current are known to be temperature-dependent and can yield this dependency, including the rate of dark avalanches, the glass resistivity and the chamber mechanical reaction to changes of internal pressure.

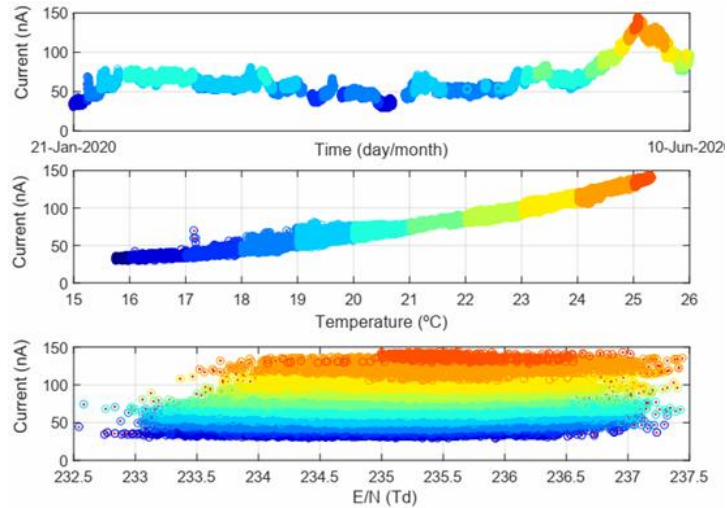

**Figure 4:** Effect of the temperature on the background current at constant reduced electric field (E/N). The current is very well correlated with temperature within the 10°C range covered in the measurements (middle panel) without evidence of a secondary dependence on the time elapsed, indicating an invariant gas quality inside the gap after 6 months sealed.

### 3.3 Preliminary charge and time resolution measurements

While the first sealed chamber was used for continuous monitoring of the current at "constant" E/N, the other one was instrumented to measure the charge spectrum, count rate and time resolution. These are preliminary measurements done in one pad only. For time measurements was required a coincidence between the pad and a small 40x40 mm2 area, sniffer 1 mm single gap RPC. The sniffer RPC operates with pure Tetrafluoroethane in open loop. The uppermost plot of Figure 5 shows the charge spectra in self-trigger under constant irradiation from a 60Co source. The high voltage is increased from 5.6 kV to 6.4 kV in steps of 100 Volts. As expected in a healthy chamber, the average charge increases with the voltage (the spectra are shifted towards higher charges). The very low streamer fraction could suggest some chamber saturation, which is something we have already observed in similar gaps at open loop. In the middle plot of Figure 5 the irradiation rate was varied, showing a slight decrease of the average charge for the higher rates, which is an expected effect owing to the resistivity of the glass electrodes. The lowermost plot shows the time measurements with muons. The time was strongly correlated with charge and the data in the plot was already corrected from that effect. The observed minimum of 508 ps at 245 Td is a good value for 1 mm gap chamber. For comparison was added the data for the sniffer chamber taken with 511 keV photons at very low counting rates, where the lower value is around 400 ps with pure Tetrafluoroethane.

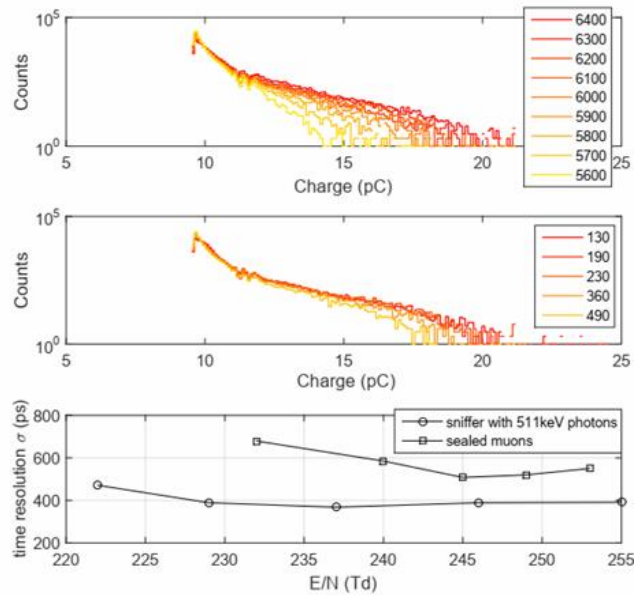

**Figure 5:** In the uppermost plot it is shown the charge spectra in self-trigger for a fixed irradiation with $^{60}$Co for different values of high voltage. The average charge increases with the voltage, as expected. In the middle plot the gain was fixed and the irradiation increased, from 130 to 490 Hz, decreasing the average charge as the irradiation increases. In the bottom plot is shown the time resolution corrected from charge as function of the reduced electric field, with very satisfactory values for this type of chamber.

## 4. Conclusion

We have built and tested two 1 mm gap RPCs with area of 40x40cm2 and a special construction aimed at extremely low gas flow operation. Following initial cleanup and filling with Tetrafluorethane, after six months of operation at zero gas flow and only occasional external irradiation we could not find any evidence that the behaviour of the chamber was influenced by the time elapsed, suggesting a stable gas quality in the gap. Measurements of the time resolution and self-trigger charge spectra indicated a normal operation as far as it can be perceived by these quantities. These results allow considering that permanently sealed RPCs may be within reach, with obvious practical and environmental advantages.

**5. Future work**

It is important to continue monitoring these two chambers since they are the best references available for time endurance. New chambers should be assembled considering all the important issues for a perfect characterization of an RPC. Finally, some physics should be done with them, most probably a muon telescope.
Finding a way to apply this concept for timing is also an open question and some efforts should be invested here as well.

**Acknowledgments**

This work is supported by Portuguese national funds OE/FCT, Lisboa2020, Compete2020, Portugal 2020, FEDER; OE, FCT-Portugal, CERN/FIS-PAR/0023/2017 and CERN/FIS-PAR/0034/2019